\pdfoutput=1

\documentclass[a4paper]{article}
\usepackage[dvipsnames]{xcolor}
\usepackage{caption}
\usepackage{subcaption}
\usepackage{INTERSPEECH2020}
\usepackage{tikz}
\usepackage{pgfplots}
\usepgfplotslibrary{units}
\usetikzlibrary{spy,backgrounds}
\usepackage{pgfplotstable}
\pgfplotsset{compat=1.14}
\usepackage{float}
\usepackage{url}

\usepackage{balance}

\title{Speech Pseudonymisation Assessment Using Voice Similarity Matrices}
\name{Paul-Gauthier No\'e$^1$, Jean-Fran\c{c}ois Bonastre$^1$, Driss Matrouf$^1$, Natalia Tomashenko$^1$,\\ Andreas Nautsch$^2$ and Nicholas Evans$^2$}

\address{
  $^1$Laboratoire Informatique d'Avignon (LIA), Avignon Universit\'e, France\\
  $^2$Digital Security Department, EURECOM, France}
\email{paul-gauthier.noe@univ-avignon.fr}

\begin{document}

\maketitle
\begin{abstract}
The proliferation of speech technologies and rising privacy legislation calls for the development of privacy preservation solutions for speech applications. These are essential since speech signals convey a wealth of rich, personal and potentially sensitive information.  Anonymisation, the focus of the recent VoicePrivacy initiative, is one strategy to protect speaker identity information.
Pseudonymisation solutions aim not only to mask the speaker identity and preserve the linguistic content, quality and naturalness, as is the goal of anonymisation, but also to preserve voice distinctiveness.  Existing metrics for the assessment of anonymisation are ill-suited and those for the assessment of pseudonymisation are completely lacking. Based upon voice similarity matrices, this paper proposes the first intuitive visualisation of pseudonymisation performance for speech signals and two novel metrics for objective assessment.  They reflect the two, key pseudonymisation requirements of de-identification and voice distinctiveness.
\end{abstract}

\section{Introduction}
The ubiquity and proliferation of speech technologies and the increase in data protection regulation such as the European General Data Protection Regulation (GDPR)~\cite{EU-GDPR-2016} has fueled interests in privacy preservation solutions for speech data~\cite{NAUTSCH2019441}. There are two general strategies: encryption and anonymisation.  Encryption is applied to protect speech data from interception and eavesdropping.  Anonymisation
aims to ensure that the protected speech data cannot be linked to the original speaker.

With very few solutions having been proposed, and with the few existing solutions achieving only modest levels of anonymisation, the VoicePrivacy initiative\footnote{\url{https://voiceprivacychallenge.org}}~\cite{tomashenko2020voiceprivacy} was launched in 2019 to promote the consideration of privacy and to foster progress in privacy preservation.  VoicePrivacy takes the form of a challenge in which participants are tasked with the development of anonymisation solutions to (i)~suppress 
(as much as possible) the
speaker identity from an utterance while nonetheless preserving voice distinctiveness and (ii)~leave intact 
(as much as possible) 
the linguistic content, quality and naturalness.
The requirement for voice distinctiveness implies that anonymised voices remain distinguishable and that all utterances from the same original speaker are anonymised with the same \emph{pseudovoice}. Such a requirement avoids confusion between speakers during a dialogue session and thus allows speaker diarization. We hence refer to the process to meet all these requirements as \emph{pseudonymisation}.

While VoicePrivacy stands to make substantial inroads, 
it is clear that the metrics used to assess pseudonymisation performance are far from being straightforward; they must reflect multifaceted criteria.  The work in this paper is concerned with metrics that reflect criteria related exclusively to the speaker \emph{identity}; it is not concerned with complementary metrics for assessing the preservation of linguistic content etc.  Most of the prior work including VoicePrivacy, e.g.~\cite{tomashenko2020voiceprivacy,fang2019speaker,srivastava2019evaluating}, measures privacy using trivial, generic metrics such as the Equal Error Rate (EER) estimated from Automatic Speaker Verification (ASV) experiments.  The general idea is to gauge performance by comparing the EER using original speech data to that obtained using speech data after de-identification; the greater the difference, or the higher the EER, the better the de-identification and privacy. 

Despite its simplicity and ease of interpretablity, the EER is ill-suited as a measure of privacy.  Principally, this is because the EER reflects the perspectives of an evaluator and not those of a \emph{privacy adversary}.  While a framework to overcome these issues is proposed in~\cite{nautsch2020}, it addresses only one component of the pseudonymisation problem, namely that relating specifically to de-identification; it does not reflect \emph{voice distinctiveness}. A solution to address both, i.e.\ a solution for the assessment of pseudonymisation, is the novel contribution in this paper.

We propose two pseudonymisation metrics for the assessment of de-identification and of voice distinctiveness.  Voice similarity matrices, upon which the two objective metrics are inspired, provide easily-interpretable visualisations of any speaker-dependent pseudonymisation behaviour and performance.  First, with a widely established privacy preservation terminology currently lacking, we provide definitions of de-identification and voice distinctiveness.  We then present voice similarity matrices and show how the two metrics are derived from them. Finally, we present the results of pseudonymisation experiments performed using the VoicePrivacy 2020 data sets and baseline systems.

\section{Pseudonymisation}

Many of the terms used in privacy research are ill-defined or at least lack a shared understanding within the speech community.  We define here more precisely the two requirements for pseudonymisation, how they relate to other terms referred to in the literature and how our work relates to them.  The requirements are as follows.

\begin{itemize}

    \item \textbf{De-identification:} a process to conceal in a speech utterance the true speaker identity~\cite{7285021,jin2009speaker,Bahmaninezhad2018}, also referred to as speaker identity masking~\cite{6859761} or voice disguise~\cite{Hautamaki2018,6423510}.
 
    \item \textbf{Voice distinctiveness:} de-identified voices should remain distinguishable within one session (e.g.~a single teleconference) such that different speakers still have different, but consistent voices, i.e.~protected utterances produced by the same speaker should be mutually linkable within a session, but they should not be linkable to the original unprotected voice. Moreover, pseudovoices should not be linkable between pseudonymised sessions but this aspect is not assessed in this work.
    Voice distinctiveness is different to \emph{voice-indistinguishability}, a term coined in~\cite{han2020voice}. The latter refers specifically and only to the unlinkability between original and protected voices.

\end{itemize}

Both anonymisation and pseudonymisation are defined within the European GDPR~\cite{EU-GDPR-2016}. The GDPR specifies that anonymisation should be irreversible whereas pseudonymisation involves the replacement of an identity with a pseudo-identity. Thus in our speech pseudonymisation framework, the mapping between unprotected and protected voices should be injective (one-to-one mapping) in order to produce distinct pseudovoices. Hence, the pseudonymisation mapping may be reversible (at least in a single session)
which is incompatible with the irreversibility requirement for anonymisation.

This paper proposes visualisations and metrics to assess the \emph{level} of de-identification, i.e.~the uncertainty in the linkability between a given utterance and the speaker identity, and the \emph{level} to which voice distinctiveness is altered in the protected space. Each speaker should have their \emph{own} protected voice.  In terms of established speech research terms, speaker diarization should perform similarly in both unprotected and protected domains. 


\section{Voice Similarity Matrices, a Visualisation}

This section describes voice similarity matrices for the assessment of pseudonymisation according to the two requirements of de-identification and voice distinctiveness. These matrices are similar to conventional confusion matrices except that they are formed with classification \emph{scores} resulting from the exhaustive comparison of utterances collected from a set of speakers.  Scores take the form of posterior probabilities, a visualisation of which is provided in the form of a heatmap.

\subsection{Voice Similarity Matrix}
\label{sec:matrices}

Let $lr(x,y)$ denote the likelihood-ratio score from the comparison of two speech segments $x$ and $y$. Assuming equal priors, it is expressed in terms of the posterior probabilities:
\begin{equation}
    lr(x,y) = \frac{P(H_{\text{tar}}|x,y)}{P(H_{\text{imp}}|x,y)}
\end{equation}
where $H_\text{tar}$ is the target proposition ($x$ and $y$ were uttered by the same speaker) and where $H_\text{imp}$ is the complementary impostor proposition. Scores, usually in the form of the log-likelihood-ratio ($llr$), can be calibrated~\cite{pav_brummer} in order to produce so-called \emph{oracle} scores.  The latter are used to compute the \emph{voice similarity} which, for two speakers $i$ and $j$, we define as:
\begin{equation}
    S(i,j) = \mathrm{sigmoid}\left({\displaystyle\sum_{\substack{1 \le k \le n_{i}\\1 \le l \le n_{j}}}\frac{llr(x^{(i)}_{k},x^{(j)}_{l})}{n_{i}n_{j} }}\right)
    \label{equ::S}
\end{equation}
which represents the posterior of the averaged $llr$, where $x^{(p)}_{q}$ is the $q$-th segment of the $p$-th speaker, $n_{p}$ is the number of segments from the $p$-th speaker and 
$\mathrm{sigmoid}(y)=({1+\exp^{-y}})^{-1}$. For $i=j$, scores for which $k=l$ are removed from the average in~(\ref{equ::S}) in order to avoid the consideration of identical speech segments which could lead to an overestimated similarity. While the average in~(\ref{equ::S}) operates upon log-likelihood-ratios, use of the $\mathrm{sigmoid}$ function yields voice similarity scores in posterior probability space. The voice similarity matrix $M$ is then given by $M=(S(i,j))_{1 \le i \le N,1 \le j \le N}$
where $N$ is the number of speakers. An example voice similarity matrix is illustrated in the \textbf{top-left quadrant} of Fig.~\ref{subfig:nonindeal1}.  The horizontal (left-to-right) and vertical axes (top-to-bottom) indicate the speaker indices $i$ and $j$ for $N=10$ speakers.  In this example, the diagonal elements depict the dominant average similarity between same-speaker utterances, i.e.~target trial comparisons that result in a higher $S(i,j)$.  The off-diagonal elements depict lower average similarity between different speakers, i.e.~impostor trial comparisons that result in a lower $S(i,j)$.  

\subsection{Visualisation of pseudonymisation performance}

We now explain how voice similarity matrices are used to assess pseudonymisation systems.  Pseudonymisation is applied to transform a set of original speech segments (O) to a set of protected, pseudonymised speech segments (P). We then define four voice similarity matrices. $M_{\text{OO}}$ and $M_{\text{PP}}$ reflect voice similarity \emph{within} the original and pseudonymised speech segment sets. The other two, $M_{\text{OP}}$ and $M_{\text{PO}}$, reflect the voice similarity \emph{between} the original and the pseudonymised sets. $M_{\text{OP}}=(M_{\text{PO}})^{\top}$ where $(\cdot)^{\top}$ denotes the transpose operator. However, as the voice similarity $S$ is assumed to be symmetric, all matrices are symmetric and $M_{\text{OP}}=M_{\text{PO}}$.
In the remainder of this paper, we hence refer only to $M_{\text{OO}}$, $M_{\text{OP}}$ and $M_{\text{PP}}$.

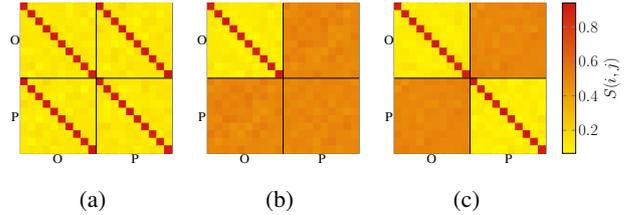
\begin{figure}
\begin{minipage}{0.14\textwidth}
\begin{subfigure}[b]{0.2\textwidth}
\begin{tikzpicture}[scale=0.35]
  \begin{axis}[view={0}{90},
            colormap={slategraywhite}{
            rgb255=(255,247,0)
            rgb255=(209,20,20)
            },
            axis equal image,
            yticklabels={,,},
            xticklabels={,,},
            enlargelimits=false,
            ]
    \addplot [matrix plot*, point meta=explicit] file {matrices/particular/non_ideal1.mat.txt};
    \draw [line width=0.2mm] (9.5,0) -- (9.5,20.5);
    \draw [line width=0.2mm] (-0.5,10.5) -- (20,10.5);
  \end{axis}
  \node at (1.4,-0.2) {\tiny{O}};
  \node at (4.3,-0.2) {\tiny{P}};
  \node at (-0.2,4.3) {\tiny{O}};
  \node at (-0.2,1.4) {\tiny{P}};
\end{tikzpicture}
\end{subfigure}
\subcaption{}
\label{subfig:nonindeal1}
\end{minipage}
\begin{minipage}{0.14\textwidth}
\begin{subfigure}[b]{0.2\textwidth}
\begin{tikzpicture}[scale=0.35]
  \begin{axis}[view={0}{90},
            colormap={slategraywhite}{
            rgb255=(255,247,0)
            rgb255=(209,20,20)
            },
            axis equal image,
            yticklabels={,,},
            xticklabels={,,},
            enlargelimits=false,
            ]
    \addplot [matrix plot*, point meta=explicit] file {matrices/particular/non_ideal2.mat.txt};
    \draw [line width=0.2mm] (9.5,0) -- (9.5,20.5);
    \draw [line width=0.2mm] (-0.5,10.5) -- (20,10.5);
  \end{axis}
  \node at (1.4,-0.2) {\tiny{O}};
  \node at (4.3,-0.2) {\tiny{P}};
  \node at (-0.2,4.3) {\tiny{O}};
  \node at (-0.2,1.4) {\tiny{P}};
\end{tikzpicture}
\end{subfigure}
\subcaption{}
\label{subfig:nonindeal2}
\end{minipage}
\begin{minipage}{0.14\textwidth}
\begin{subfigure}[b]{0.2\textwidth}
\begin{tikzpicture}[scale=0.35]
  \begin{axis}[view={0}{90},
            colorbar,
            colorbar style={
                xshift=1em,
                ylabel={$S(i,j)$},
                font=\LARGE
            },
            colormap={slategraywhite}{
            rgb255=(255,247,0)
            rgb255=(209,20,20)
            },
            axis equal image,
            yticklabels={,,},
            xticklabels={,,},
            enlargelimits=false,
            ]
    \addplot [matrix plot*, point meta=explicit] file {matrices/particular/ideal.mat.txt};
    \draw [line width=0.2mm] (9.5,0) -- (9.5,20.5);
    \draw [line width=0.2mm] (-0.5,10.5) -- (20,10.5);
  \end{axis}
  \node at (1.4,-0.2) {\tiny{O}};
  \node at (4.3,-0.2) {\tiny{P}};
  \node at (-0.2,4.3) {\tiny{O}};
  \node at (-0.2,1.4) {\tiny{P}};
\end{tikzpicture}
\end{subfigure}
\subcaption{}
\label{subfig:ideal}
\end{minipage}
\caption{Three artificial similarity matrices. The upper-left matrix is $M_{\text{OO}}$, the upper-right and lower-left are $M_{\text{OP}}$ whereas the lower-right is $M_{\text{PP}}$.}
\label{fig:examples_matrices}
\end{figure}

Fig.~\ref{fig:examples_matrices} shows three example similarity matrices.  In each case, $M_{\text{OO}}$ is in the upper-left quadrant, $M_{\text{OP}}$ is in the upper-right quadrant and $M_{\text{PP}}$ is in the lower-right quadrant. Fig.~\ref{subfig:nonindeal1} illustrates the impact upon voice similarity of a poor pseudonymisation system; voice similarities between original, pseudonymised and original-pseudonymised segments are more-or-less identical.  Pseudomynisation achieves nothing, even if voice distinctiveness is preserved ($M_{\text{PP}}$ still exhibits a dominant diagonal). Fig.~\ref{subfig:nonindeal2} illustrates the behaviour of a different pseudonymisation system for which voice distinctiveness is lost (there is no dominant diagonal in $M_{\text{PP}}$). This system, however, is more successful in de-identification ($M_{\text{OP}}$ also has no dominant diagonal). Fig.~\ref{subfig:ideal} visualises the performance of an ideal case in which both de-identification and voice distinctiveness criteria are met: $M_{\text{OP}}$ is uniform without a dominant diagonal; $M_{\text{PP}}$ does exhibit a dominant diagonal.

The three $M$ matrices serve to visualise any differences in pseudonymisation performance at the speaker level. While these are not apparent in the artificial examples in Fig.~\ref{fig:examples_matrices}, since ASV performance typically varies across different speakers~\cite{Doddington1998SHEEPGL,kahn2010}, they are expected in practice for real speech data (see Section \ref{sec:results}).
We show next how the visualisations shown in Fig.~\ref{fig:examples_matrices} can be used to derive objective measures of both de-identification and voice distinctiveness.

\section{Proposed Metrics}
\label{sec:metrics}

$M_{\text{OO}}$ and $M_{\text{PP}}$ show voice distinctiveness in original and pseudonymised space respectively, while $M_{\text{OP}}$ shows the ease with which speakers in original space can be linked to speakers in pseudonymised space (and vice versa).  This information is most easily visualised by the presence or absence of a dominant diagonal.  
Accordingly, a measure of de-identification and voice distinctiveness can be captured by quantification of diagonal dominance: the key idea behind both proposed metrics.

For any of the three $M$ matrices, the diagonal dominance $D_\text{diag}(M)$ is defined as the absolute difference between the averages of the diagonal and the off-diagonal elements:
\begin{equation}
    D_{\text{diag}}(M)\hspace{-0.7mm}=\hspace{-0.7mm}\displaystyle
    \Big|
    \Big(
    \sum_{1\leq i \leq N} \frac{S(i,i)}{N}
    \Big)
    \displaystyle
    - 
    \Big(
    \sum_{\substack{1 \le j \le N\\1 \le k \le N\\j \neq k}}
    \frac{S(j,k)}{N(N-1)}
    \Big)
    \Big|
    \label{eq:ddiag}
\end{equation}
$D_\text{diag}$ will be $0$ for a constant/uniform matrix and $1$ for an identity matrix as well as for a matrix where all diagonal elements are $0$ and all off-diagonal elements are $1$.

\subsection{De-identification}

A measure of de-identification performance is obtained from the comparison of $D_{\text{diag}}$ in the original space to that \emph{between} the original and pseudonymised space. Assuming that $D_\text{diag}(M_{\text{OO}})$ is strictly positive and that $D_\text{diag}(M_{\text{OP}}) \le D_\text{diag}(M_{\text{OO}})$ (de-identification should always reduce the diagonal dominance in $M_{\text{OP}}$), then de-identification performance is measured according to:
\begin{equation}
    \mathrm{DeID}= 1-\frac{D_\text{diag}(M_{\text{OP}})}{D_\text{diag}(M_{\text{OO}})}
\end{equation}
The denominator acts to normalise $D_{\text{diag}}(M_{\text{OP}})$ so that a pseudonymisation solution that does nothing will yield a $\mathrm{DeID}$ of 0\%.  Conversely, if the de-identification is optimal, i.e.\ $D_\text{diag}(M_{\text{OP}})=0$, then $\mathrm{DeID}=100\%$.
    
\subsection{Voice distinctiveness}

Pseudonymisation can both degrade or improve voice distinctiveness. Motivated from electrical engineering and signal processing, we report a gain value on a decibel (dB) scale as follows:
\begin{equation}
    G_{\text{VD}} = 10\log_{10} \Big( \frac{D_\text{diag}(M_{\text{PP}})}{D_\text{diag}(M_{\text{\text{OO}}})}\Big)
\end{equation}

Gains above 0~dB indicate an increase in voice distinctiveness. Gains below 0~dB indicate a degradation whereas a value of exactly 0~dB indicates that voice distinctiveness in original space is preserved in pseudonymised space. 

\section{Pseudonymisation, a Case Study}

In this section we present an analysis of pseudonymisation performance using the proposed matrices and the de-identification and voice distinctiveness metrics.\footnote{The matrices and metrics are integrated in the VoicePrivacy Challenge: \url{https://github.com/Voice-Privacy-Challenge}.}

\subsection{Data sets, protocols and baselines}

This work was performed using the VoicePrivacy Challenge 2020~\cite{tomashenko2020voiceprivacy} data sets and the two associated baseline systems. The primary baseline, inspired from~\cite{fuming2019}, is based on a x-vector pooling and neural waveform model resynthesis approach. The secondary baseline is based on vocal tract filter transformations using McAdams coefficients~\cite{EURECOM+6190}.

Results are reported for the VoicePrivacy 2020 development data sets. They are drawn from \emph{LibriSpeech-dev-clean}~\cite{panayotov2015librispeech} and \emph{VCTK-dev}~\cite{yamagishi2019cstr}. We use the trial parts of the challenge development data sets. Details of both data sets and baselines can be found in~\cite{tomashenkovoiceprivacy}. For brevity, the sets are renamed as shown in Tab.~\ref{tab:rename}.

\begin{table}
\centering
    \begin{tabular}{ccc}
        \hline
        \# & Official name & Short name \\
        \hline
        1 & libri\_dev\_trials\_f & ldtf \\
        2 & libri\_dev\_trials\_m & ldtm \\
        3 & vctk\_dev\_trials\_f & vdtf \\
        4 & vctk\_dev\_trials\_m & vdtm \\
        5 & vctk\_dev\_trials\_f\_common & vdtfc \\
        6 & vctk\_dev\_trials\_m\_common & vdtmc \\
        \hline
    \end{tabular}
    \caption{Renaming of the development sets presented in~\cite{tomashenkovoiceprivacy}.}
    \label{tab:rename}
\end{table}

As per challenge conditions, scores are obtained from the comparison of x-vectors~\cite{snyder2018x} using probabilistic linear discriminant analysis~\cite{pldaIoffe}. 
Each of the three score sets used for the computation of the three similarity matrices are oracle calibrated~\cite{pav_brummer}.

\subsection{Pseudonymisation assessment results}
\label{sec:results}

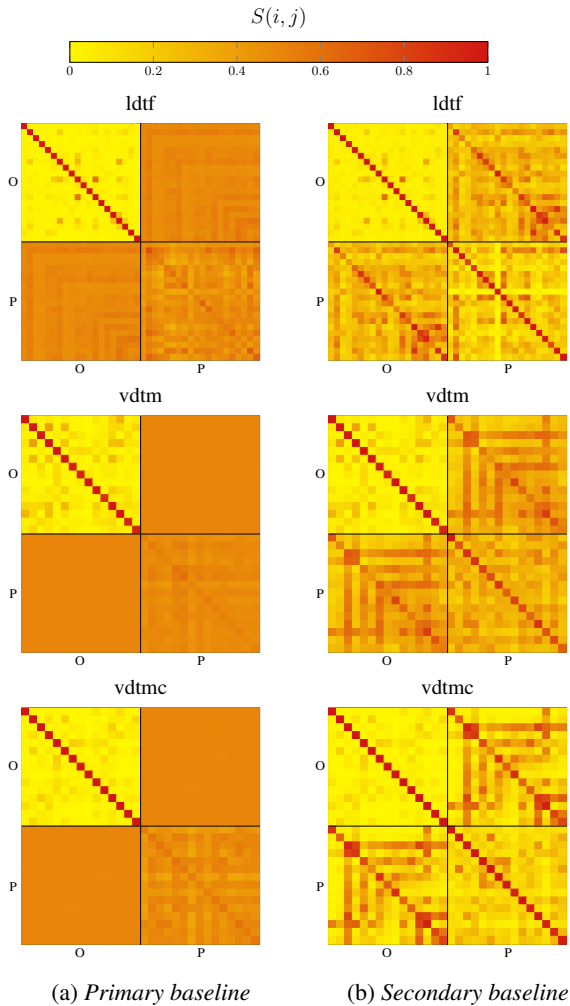
\begin{figure}[H]
\begin{minipage}{0.43\textwidth}
\centering
\begin{tikzpicture}[scale=0.55]
    \begin{axis}[view={0}{90},
            hide axis,
            scale only axis,
            height=0pt,
            width=0pt,
            colormap={slategraywhite}{
            rgb255=(255,247,0)
            rgb255=(209,20,20)
            },
            colorbar horizontal,
            point meta min=0,
            point meta max=1,
            colorbar style={
            width=10cm,
            title=\Large{$S(i,j)$}, xtick={0,0.2,0.4,...,1}
            }]
            ]
            \addplot [draw=none] coordinates {(0,0)};
    \end{axis}

\end{tikzpicture}
\end{minipage}

\begin{minipage}{0.233\textwidth}
\begin{subfigure}[b]{0.2\textwidth}
\begin{tikzpicture}[scale=0.55]
  \begin{axis}[view={0}{90},
            colormap={slategraywhite}{
            rgb255=(255,247,0)
            rgb255=(209,20,20)
            },
            axis equal image,
            yticklabels={,,},
            xticklabels={,,},
            enlargelimits=false,
            title=\Large{ldtf},
            ]
    \addplot [matrix plot*, point meta=explicit] file {matrices/final_similarity_matrices/first_baseline/ldtf.mat.txt};
    \draw [line width=0.2mm] (19.5,0) -- (19.5,41);
    \draw [line width=0.2mm] (-0.5,20.5) -- (41,20.5);
  \end{axis}
  \node at (1.4,-0.2) {\tiny{O}};
  \node at (4.3,-0.2) {\tiny{P}};
  \node at (-0.2,4.3) {\tiny{O}};
  \node at (-0.2,1.4) {\tiny{P}};
\end{tikzpicture}
\end{subfigure}

\begin{subfigure}[b]{0.2\textwidth}
\begin{tikzpicture}[scale=0.55]
  \begin{axis}[view={0}{90},
            colormap={slategraywhite}{
            rgb255=(255,247,0)
            rgb255=(209,20,20)
            },
            axis equal image,
            yticklabels={,,},
            xticklabels={,,},
            enlargelimits=false,
            title=\Large{vdtm},
            ]
    \addplot [matrix plot*, point meta=explicit] file {matrices/final_similarity_matrices/first_baseline/vdtm.mat.txt};
    \draw [line width=0.2mm] (14.5,-0.5) -- (14.5,30.5);
    \draw [line width=0.2mm] (-0.5,15.5) -- (30,15.5);
  \end{axis}
  \node at (1.4,-0.2) {\tiny{O}};
  \node at (4.3,-0.2) {\tiny{P}};
  \node at (-0.2,4.3) {\tiny{O}};
  \node at (-0.2,1.4) {\tiny{P}};
\end{tikzpicture}
\end{subfigure}

\begin{subfigure}[b]{0.2\textwidth}
\begin{tikzpicture}[scale=0.55]
  \begin{axis}[view={0}{90},
            colormap={slategraywhite}{
            rgb255=(255,247,0)
            rgb255=(209,20,20)
            },
            axis equal image,
            yticklabels={,,},
            xticklabels={,,},
            enlargelimits=false,
            title=\Large{vdtmc},
            ]
    \addplot [matrix plot*, point meta=explicit] file {matrices/final_similarity_matrices/first_baseline/vdtmc.mat.txt};
    \draw [line width=0.2mm] (14.5,-0.5) -- (14.5,30.5);
    \draw [line width=0.2mm] (-0.5,15.5) -- (30,15.5);
  \end{axis}
  \node at (1.4,-0.2) {\tiny{O}};
  \node at (4.3,-0.2) {\tiny{P}};
  \node at (-0.2,4.3) {\tiny{O}};
  \node at (-0.2,1.4) {\tiny{P}};
\end{tikzpicture}
\end{subfigure}
\subcaption{Primary baseline}
\end{minipage}
\begin{minipage}{0.233\textwidth}
\begin{subfigure}[b]{0.3\textwidth}
\begin{tikzpicture}[scale=0.55]
  \begin{axis}[view={0}{90},
            colormap={slategraywhite}{
            rgb255=(255,247,0)
            rgb255=(209,20,20)
            },
            axis equal image,
            yticklabels={,,},
            xticklabels={,,},
            enlargelimits=false,
            title=\Large{ldtf},
            ]
    \addplot [matrix plot*, point meta=explicit] file {matrices/final_similarity_matrices/second_baseline/ldtf.mat.txt};
    \draw [line width=0.2mm] (19.5,0) -- (19.5,41);
    \draw [line width=0.2mm] (-0.5,20.5) -- (41,20.5);
  \end{axis}
  \node at (1.4,-0.2) {\tiny{O}};
  \node at (4.3,-0.2) {\tiny{P}};
  \node at (-0.2,4.3) {\tiny{O}};
  \node at (-0.2,1.4) {\tiny{P}};
\end{tikzpicture}
\end{subfigure}

\begin{subfigure}[b]{0.3\textwidth}
\begin{tikzpicture}[scale=0.55]
  \begin{axis}[view={0}{90},
            colormap={slategraywhite}{
            rgb255=(255,247,0)
            rgb255=(209,20,20)
            },
            axis equal image,
            yticklabels={,,},
            xticklabels={,,},
            enlargelimits=false,
            title=\Large{vdtm},
            ]
    \addplot [matrix plot*, point meta=explicit] file {matrices/final_similarity_matrices/second_baseline/vdtm.mat.txt};
    \draw [line width=0.2mm] (14.5,-0.5) -- (14.5,30.5);
    \draw [line width=0.2mm] (-0.5,15.5) -- (30,15.5);
  \end{axis}
  \node at (1.4,-0.2) {\tiny{O}};
  \node at (4.3,-0.2) {\tiny{P}};
  \node at (-0.2,4.3) {\tiny{O}};
  \node at (-0.2,1.4) {\tiny{P}};
\end{tikzpicture}
\end{subfigure}

\begin{subfigure}[b]{0.3\textwidth}
\begin{tikzpicture}[scale=0.55]
  \begin{axis}[view={0}{90},
            colormap={slategraywhite}{
            rgb255=(255,247,0)
            rgb255=(209,20,20)
            },
            axis equal image,
            yticklabels={,,},
            xticklabels={,,},
            enlargelimits=false,
            title=\Large{vdtmc},
            ]
    \addplot [matrix plot*, point meta=explicit] file {matrices/final_similarity_matrices/second_baseline/vdtmc.mat.txt};
    \draw [line width=0.2mm] (14.5,-0.5) -- (14.5,30.5);
    \draw [line width=0.2mm] (-0.5,15.5) -- (30,15.5);
  \end{axis}
  \node at (1.4,-0.2) {\tiny{O}};
  \node at (4.3,-0.2) {\tiny{P}};
  \node at (-0.2,4.3) {\tiny{O}};
  \node at (-0.2,1.4) {\tiny{P}};
\end{tikzpicture}
\end{subfigure}
\subcaption{Secondary baseline}
\end{minipage}
\caption{Voice similarity matrices for the two baselines on ldtf, vdtm and vdtmc.}
\label{fig:matrices_baselines}
\end{figure}

Fig.~\ref{fig:matrices_baselines} provides separate visualisations of pseudonymisation performance for three of the data sets.  They show that the primary baseline (left column) delivers better de-identification performance than the secondary baseline (right column).  For the primary baseline, entries in $M_{\text{OP}}$ (upper-right and lower-left quadrants) have values close to 0.5, indicating strong de-identification.  In contrast, $M_{\text{OP}}$ matrices for the secondary baseline show perceptible diagonals, indicating weaker de-identification.  The same visualisations show that the secondary baseline better preserves voice distinctiveness; diagonals in $M_{\text{PP}}$ matrices (lower-right quadrants) are more perceptible for the secondary than the primary baseline.

Some differences in performance across speakers and data sets are also visible in Fig.~\ref{fig:matrices_baselines}. For the primary baseline, voice distinctiveness seems to be slightly better for ldtf and vdtmc data than for vdtm data (more distinctive diagonals in $M_{\text{PP}}$ matrices).  While de-identification appears to be consistent for the primary baseline, the secondary baseline appears to perform slightly better for vdtm data than for ldft and vdtmc data, albeit it still poorly. De-identification performance is also seen to depend on the speaker, e.g.\ there are visible striations in $M_{\text{OP}}$ for the secondary baseline and vdtm and vdtmc data and, to a much lesser extent, for the primary baseline and ldtf data. The resulting voice distinctiveness also depends on the speaker, e.g.\ for the secondary baseline, the pseudovoice of one speaker in ldtf is notably more distinctive than others (pure yellow row and column apart the diagonal element in $M_{\text{PP}}$).

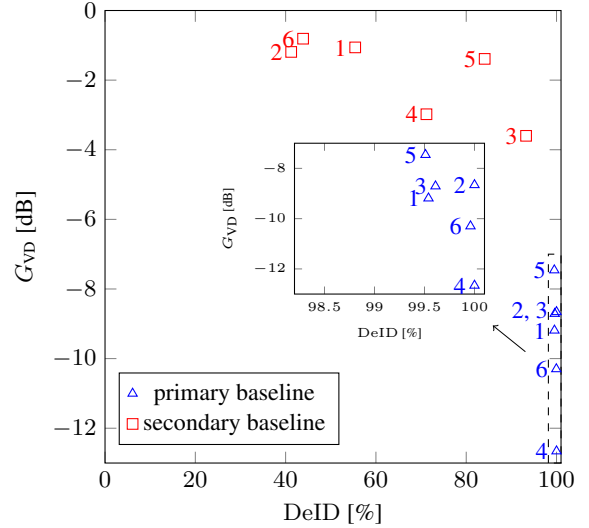
\begin{figure}
\centering
\begin{tikzpicture}
    \begin{axis}[
scale only axis=true,
width=6cm,
height=6cm,
xlabel={$\mathrm{DeID}$ [\%]},
ylabel={$G_{\text{VD}}$ [dB]},
xmin=0,
xmax=101,
ymin=-13,
ymax=0,
legend style={at={(0.03,0.12)},anchor=west},
]    
    \addplot+ [only marks,
    mark=triangle,
    mark options={draw=blue,fill=blue},
    nodes near coords={\Label},
    nodes near coords align={horizontal},
    visualization depends on={value \thisrow{set} \as \Label}
    ]
    table[x=deid, y=vu]{
    deid    vu      baseline    set
    99.54   -9.19   1           1
    100     -8.66   1           {2, 3}
    99.61   -8.71   1           ~
    100     -12.66  1           4
    99.51   -7.46   1           5
    99.96   -10.30  1           6

    
    };
    \addplot+ [only marks,
    mark=square,
    mark options={draw=red,fill=red},
    nodes near coords={\Label},
    nodes near coords align={horizontal},
    visualization depends on={value \thisrow{set} \as \Label}
    ]
    table[x=deid, y=vu, meta=baseline]{
    deid    vu      baseline    set
    55.41   -1.06   2           1
    41.23   -1.19   2           2
    93.26   -3.60   2           3
    71.19   -2.98   2           4
    84.09   -1.39   2           5
    43.87   -0.81   2           6
    };
    \legend{primary baseline,secondary baseline}
    \coordinate (pt) at (axis cs:95,-10);
    \draw [dashed] (98.2,-13) rectangle (101,-7);
    \end{axis}
    \node[pin={[pin edge={black, ->}]175:{%
        \begin{tikzpicture}[baseline,trim axis left,trim axis right]
        \begin{axis}[tiny,
        width=2.5cm,
        height=2cm,
        xlabel={$\mathrm{DeID}$ [\%]},
        ylabel={$G_{\text{VD}}$ [dB]},
        xmin=98.2,xmax=100.1,
        ymin=-13,ymax=-7,
        scale only axis=true,
        ]
        \addplot +[only marks, mark=triangle, mark options={draw=blue,fill=blue},nodes near coords={\Label}, nodes near coords align={horizontal}, visualization depends on={value \thisrow{set} \as \Label}
        ] 
        table[x=deid, y=vu, row sep=\\]{
        deid    vu      baseline    set\\
        99.54   -9.19   1           1\\
        100     -8.66   1           2\\
        99.61   -8.71   1           3\\
        100     -12.66  1           4\\
        99.51   -7.46   1           5\\
        99.96   -10.30  1           6\\
        };
        \end{axis}
        \end{tikzpicture}%
    }}] at (pt) {};
\end{tikzpicture}
\caption{Scatter plot of Gain of Voice Distinctiveness ($G_{\text{VD}}$) vs.\ De-Identification ($\mathrm{DeID}$) for both baselines and each set.}
\label{fig:scatter}
\end{figure}

$\mathrm{DeID}$ and $G_{\text{VD}}$ results for each data set and for the primary (blue triangles) and secondary (red squares) baselines are shown in Fig.~\ref{fig:scatter} and Tab.~\ref{tab:results}.  $\mathrm{DeID}$ rates for the primary baseline are consistently close to 100\%, whereas those for the secondary baseline vary between approximately 44\% and 93\%.  $G_{\text{VD}}$ rates for the secondary baseline are consistently close to zero, whereas those for the primary baseline vary between approximately -7.5 and -13~dB. Such substantial degradations to voice distinctiveness in comparison to the secondary baseline are no surprise since the primary baseline is based upon x-vector averaging over a subset of speakers.  These objective results confirm observations from the visualisations in Fig.~\ref{fig:matrices_baselines}.\\

\begin{center}
\begin{table}
\begin{tabular}{ccccc}
    \cline{2-5}
     & \multicolumn{2}{c}{primary baseline} & \multicolumn{2}{c}{secondary baseline} \\
     \hline
     set & $\mathrm{DeID}$ [\%] & $G_{\text{VD}}$ [dB] & $\mathrm{DeID}$ [\%] & $G_{\text{VD}}$ [dB]\\
     \hline
     ldtf & 99.54 & -9.19 & 55.41 & -1.06\\
     ldtm & 100 & -8.66 & 41.23 & -1.19 \\
     vdtf & 99.61 & -8.71 & 93.26 & -3.60 \\
     vdtm & 100 & -12.66 & 71.19 & -2.98\\
     vdtfc & 99.51 & -7.46 & 84.09 & -1.39\\
     vdtmc & 99.96 & -10.30 & 43.87 & -0.81 \\
    \hline
\end{tabular}
    \caption{Results of De-Identification and Gain of Voice Distinctiveness for both baselines and each set.}
\label{tab:results}
\end{table}
\end{center}

\section{Conclusions}

This paper describes an approach to visualise the de-identification and voice distinctiveness delivered by pseudonymisation solutions and defines objective metrics. Voice similarity matrices, upon which the visualisations and metrics are based, provide revealing, snapshot insights into pseudonymisation performance.  They expose differences in performance across different data and speakers.  For the latter, visualisations show that, while a particular pseudonymisation solution might perform well on average, it might leave some subjects with relatively weak protection, a finding which is not evident from results derived from objective metrics alone.  Other findings point towards a possible trade-off or compromise between de-identification and voice distinctiveness.  One pseudonymisation solution delivers near-to-perfect de-identification, whereas the other better preserves voice distinctiveness.  Solutions based upon the pooling or averaging of speaker characteristics, as it is the case for the primary baseline, may lead to losses in voice distinctiveness. 

Future work should hence investigate injective voice mapping techniques to preserve distinctiveness.
However, they will require careful design since they may jeopardise irreversibility (voices cannot be re-identified through an inverse transformation), a key requirement for anonymisation.  
A compromise solution might be to insure the injectivity for preserving the voice distinctiveness within a diaglogue session whereas adding non-injectivity or randomness between the dialogue sessions. Thus the intra-session mapping could be reversible while the inter-session mapping could be irreversible. In this case a privacy adversary would not be able to use data across sessions. Hence, metrics to the assessment of voice de-identification versus voice distinctiveness in multi-session must be elaborated in future research.

\section{Acknowledgements}

This work was supported by the JST-ANR Japanese-French project VoicePersonae.

\bibliographystyle{IEEEtran}
\balance\bibliography{mybib}

\end{document}